\documentclass[letterpaper, 10 pt, conference]{ieeeconf}  

\IEEEoverridecommandlockouts

\usepackage{graphicx}
\usepackage{multirow}
\usepackage{listings}
\usepackage{subfig}
\usepackage{amsmath, amssymb,amsfonts}
\usepackage{hyperref}
\usepackage[short]{optidef}

\title{\LARGE \bf
Real-Time Model Predictive Control for the Swing-Up Problem of an Underactuated Double Pendulum
}

\author{Blanka Burchard$^{1,2}$ and Franek Stark$^{1}$
\thanks{$^{1}$ Robotics Innovation Center, German Research Center for Artificial Intelligence (DFKI GmbH), Bremen, Germany.}%
\thanks{$^{2}$ Faculty of Electrical Engineering and Informatics, City University of applied Sciences Bremen}
\thanks{Corresponding author: blanka.burchard@dfki.de}
}

\begin{document}

\maketitle
\thispagestyle{empty}
\pagestyle{empty}

\begin{abstract}

The 3rd "AI Olympics with RealAIGym" competition poses the challenge of developing a global policy that can swing up and stabilize an underactuated 2-link system \textit{Acrobot} and/or \textit{Pendubot} from any configuration in the state space. This paper presents an optimal control-based approach using a real-time Nonlinear Model Predictive Control (MPC). The results show that the controller achieves good performance and robustness and can reliably handle disturbances. 

\end{abstract}

\section{Introduction}

The AI Olympics Competition Series held by the German Research Center for Artificial Intelligence (DFKI) and Chalmers University of Technology aims to improve robots' physical or athletic intelligence. Following the inaugural run 2023 \cite{c6} and the 2nd edition 2024 \cite{c11} which focused on benchmarking various learning and control methods for the swing-up problem of the double pendulum using one active joint and achieving robustness to disturbances, the 3rd AI Olympics competitions goal is to develop a global policy that can stabilize the system from any configuration in the state space. 
The competition evaluates controller performance for two different configurations of the double pendulum \cite{c10}: 
The Acrobot, with an actuator at the elbow, and the Pendubot, with an actuator at the shoulder. 
While machine Learning methods and particularly model-free reinforcement learning algorithms are often used in this context \cite{c3, c4}, this work aims to prove optimal control-based approaches as a competitive alternative. 

Model Predictive Control (MPC) has become a popular method in embedded control thanks to its predictive nature, robustness to uncertainty and changing conditions, and its ability to handle nonlinearities and constraints. 
Nonetheless, it also poses some challenges for real-time use due to its computational cost. Thus the proposed controller uses \textit{acados}, a framework providing fast and computationally efficient solvers for nonlinear optimal control \cite{c7}.
MPC is based on the solution of an optimal control problem (OCP) at every sampling instance, multiple numerical methods to solve OCPs can be used two of which are focused on in this paper.
 Sequential Quadratic Programming (SQP) belongs to the family of Newton-type methods and has become standard along numerical methods for solving NLP's. Furthermore, Newton-type methods hold essential features that can be exploited for real-time use as will be concretized later.  
 Differential Dynamic Programming (DDP) uses the recursive equation of dynamic programming to iteratively search for an improved trajectory \cite{c5}. Though being less effective in dealing with constraints, some works report essential speedups in certain nonlinear problems, compared to Newton-type methods \cite{c10}.

This paper is organized as follows: In \autoref{sec:problem} the discrete-time, finite-horizon Optimization Problem is formulated and the model is described. In \autoref{sec:method} the used Method, thus the high-level implementation of the controller is described. Finally, \autoref{sec:results} presents the results for both Acrobot and the Pendubot individually.

\section{Problem formulation}\label{sec:problem}
Given a nonlinear dynamical system \(\dot{x} =f(x,u)\), MPC computes the optimal control input \(u\) that minimizes a cost function $l$ over a finite, receding horizon by solving a constrained optimal control problem (OCP) at each time step. The first computed input is applied, disregarding the following ones, he state is measured and the OCP is solved again based on the new state measurement. 
A detailed overview on Model Predictive Control, its theory, and the numerical algorithms associated can be taken from \cite{c1}.
The presented approach uses a multiple shooting parametrization, thus the Optimal Control Problem is approximated by discretizing the system into \(N\) steps of step size \(\delta t\) over a time horizon \(T\), such that an NLP formulation can be stated as:
\begin{mini}{x,u}
{l_f(x[N]) + \sum_{n=0}^{N-1}
        l(x[n], u[n])}{}{}
\addConstraint{\dot{x}[n]}{= f(x[n], u[n]), \quad}{ n=0,\dots,N-1}
\addConstraint{\underline{u}}{\leq D u[n] \leq \overline{u}, \quad}{ n=0,\dots,N-1}
\addConstraint{\underline{x}}{\leq D x[n] \leq \overline{x}, \quad }{n=0,\dots,N-1}
\addConstraint{\underline{x}_f}{\leq D_f x[N] \leq \overline{x}_f}{}
\addConstraint{x_0}{=x_0}{}
\end{mini}

With $x_0$ being the measured state, $l$ the running cost function, $l_f$ the final cost function, $D$, $C$ and $D_f$ a linear constraint mapping of state and input to their lower ($\underline{x}, \underline{u}, \underline{x}_f$) and upper bound ($\overline{x}, \overline{u}, \overline{x}_f$). $x[n]$ and $u[n]$ denote state or input at the discretisation step $n$ respectively.

\subsection{Model}
The Double Pendulum is a two-link robotic arm in the vertical plane.
The state of the double pendulum is defined as $x = [q, \dot{q}]$ where $q$ comprises the two joint angles $q=[\theta_1, \theta_2]$. The Equations of motion for the double Pendulum model can be derived using the method of Lagrange and are given by:
\begin{equation}
    M(q)\ddot{q} + C(q,\dot{q})\dot{q} =\tau_g(q) + Bu
\end{equation}
where \(M(q)\) denotes the mass-inertia matrix, \(C(q, \dot{q})\) denotes the Coriolis and centrifugal matrix, \(\tau_g (q)\) comprises the gravity effects, the actuation matrix $B$ selects which actuator is active and $u$ is the input motor torque. 
A time-varying linearization of the system leads to the standard linear state-space form:
\begin{equation}
\dot{x}= 
\begin{bmatrix}
    \dot{q}\\
    M^{-1}(q)[\tau_g(q)+B(q)u-C(q,\dot{q}\dot{q})]
\end{bmatrix}
\end{equation}
The derivation of the model is based on the underactuated robotics course taught at MIT \cite{c8} and is analogous to the double pendulum model used in the RealAiGym project \cite{c9} which can be referred to for more details.

\subsection{Cost and Constraints}
The cost function is either defined as an L2-norm cost function:
\begin{align}\label{eq:cost_lin}
\begin{split}
L(x[n],u[n]) =& \frac{1}{2}\|(x[n]-\hat{x})^TQ(x([n]-\hat{x}) \\ 
&+(u[n])^TR~u[n]\|^2 \\
L_f(x,u) =& \frac{1}{2}\|(x[N]-\hat{x})^TQ_f(x[N]-\hat{x})\|^2
\end{split}
\end{align}
with $\hat{x}$ being the target state and $Q$ and $R$ the weight matrices that can be configured on controller setup.

Alternatively, the cost can be expressed as a non-linear least squares function. For that in \eqref{eq:cost_lin} the $x[n]$ and $\hat{x}$  are replaced by a non-linear function $l(x[n])$ and $l(\hat{x})$.

\section{Method}\label{sec:method}

This section illustrates the high-level implementation of the controller. Multiple options with strategies in mind for increasing robustness and real-time capabilities of the controller were implemented, some of which are described in detail below. Table \ref{table:classParameters} shows all the available options that can be configured after it's creation.

For comparison, different solving techniques provided by acados were integrated in the controller and can be used to solve the nonlinear OCP in acados including SQP, SQP for real time use (SQP-RTI) and DDP. 
SQP-RTI performs single iteration of the SQP solver and exploits high controller frequencies and the fact that steps of the SQP algorithm, notably those with the most computational burden, can be performed without knowledge of the state x, thus enabling to separate the algorithm into a preparation and a feedback phase. Computations of the preparation phase that precalculates information as far as possible without knowing the state x, are carried out asynchronously and in parallel to forward simulating the pendulum system. The feedback phase finishes the computation when a measured state is available \cite{c2}.

The fact that the angle of the joints always lies between $-\pi$ and $\pi$ was initially not reflected in the cost function which allowed for an endlessly high distance between the final state $xf$ and the current state $x$, leading to wrong solutions and nonfeasible problems becoming successively more on the long run. 

To overcome this problem, the option to embed the pendulum angles $\theta_1, \theta_2$ into $\mathcal{R}^2$ using the mapping $\theta \mapsto \left(\cos(\theta), \sin(\theta)\right)$, allowing us to compute a cost that is invariant to $2\pi$-shifts.

In an attempt to increase system stability, the option was added to let a PID Controller adjust u based on the optimal next state calculated by the OCP Solver, instead of directly applying the u of the optimized input trajectory.

A simple fallback strategy to handle infeasibility of the NLP was implemented, in which  the optimized trajectories are stored until the next solver run returns a feasible solution. Any time no solution could be constructed, the next u of the optimized input trajectory is applied and the trajectory shifted to the left by 1, removing the first element and appending 0 to the end.

The Multiple shooting grid of equally sized timesteps $\delta t$  was redefined as a non uniform grid with the time between shooting nodes $n \in [0,1....N-1]$ growing linearly dependent on $n$.

\section{Results and Discussion}\label{sec:results}
Table \ref{table:options} shows the controller options used in both Setups. According to the Hardware restrictions of the real system a maximum velocity of 30 rad/s and maximum torque of 6 Nm were used. Lastly The Cost matrices were set to 
\begin{align}
    \begin{split}
        Q =& \begin{bmatrix}
            100& 10\\
            100& 10
        \end{bmatrix},
        R = \begin{bmatrix}
            0.000001 \\
            0.000001
        \end{bmatrix} \\
        Q_f =& \begin{bmatrix}
            10000& 100\\
            10000& 100
        \end{bmatrix}
    \end{split}
\end{align}
with $Q_f$ posing a soft constraint on the final state $x_f$.
To evaluate the controller, performance and robustness scores were measured with benchmark classes provided in the RealAiGym Project. 
The results indicate that the controller achieves very good performance and robustness with both Pendubot and Acrobot, though slightly less reliable with the latter.

\subsection{Pendubot}

\begin{figure}[thpb]
  \centering
  \includegraphics[scale=0.3]{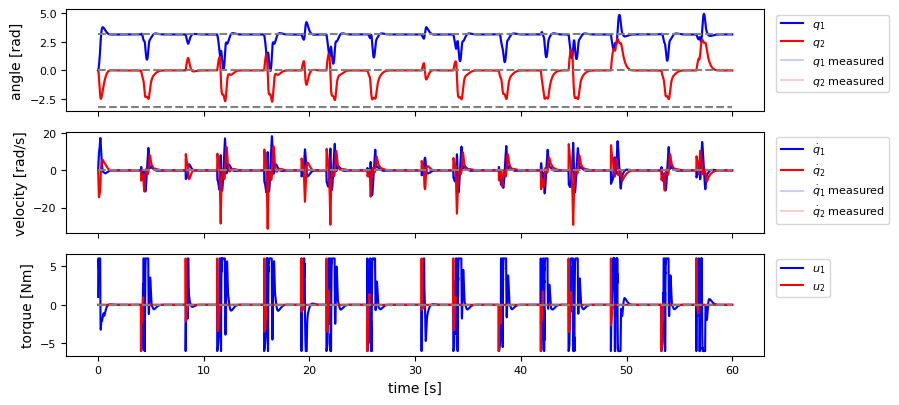}
  \caption{Simulation of the disturbed Pendubot system, controlled by the MPC}
  \label{fig:pendubot_traj}
\end{figure}

When used in the Pendubot setup the controller achieved a performance score of 0.769 with a total uptime of 46.176 seconds. The robustness metrics can be seen in Fig.~\ref{fig:robustness}.  Despite being the weakest criteria, the time delay shows quite good results up until 0.025 seconds, which has positive implications with regard to potential experiments on the real system. Noticeably the controller for the Pendubot shows weaknesses against variatons of model parameters related to the second inactive joint, particularly friction and damping. The highly non linear term for the coulomb fiction of the model turned out to have a significant effect on the stabillity of the system and can become especially challenging in the real-robot stage, thus remains a point open for future improvements.
Nontheless the controller managed to handle disturbances reliably and always managed to swing the pendulum back up in a relatively quick time as can be seen in Fig.~\ref{fig:pendubot_traj}. The graph also depicts how the controller uses the dynamics of the pendulum System to get to the goal position. After being disturbed the pendulum is accelerated into the direction of the disturbance until a bit further than the systems fixed point at $[\pi, \pi]$ . Then it is accelerated into the oposite direction, the force applied is now acting in parallel to the resetting force and uses it to swing back up.

\subsection{Acrobot}

\begin{figure}[thpb]
    \centering
    \subfloat[\centering Acrobot]{{\includegraphics[width=3.8cm]{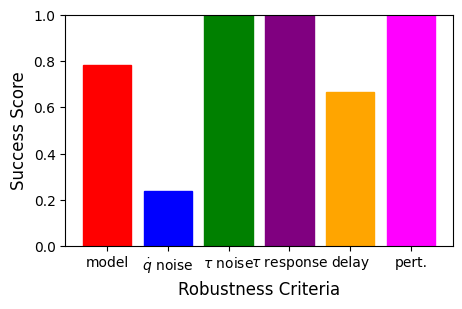} }}%
    \qquad
    \subfloat[\centering Pendubot]{{\includegraphics[width=3.8cm]{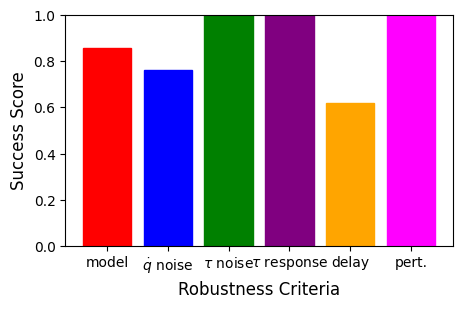} }}%
    \caption{Robustness metrics in both Setups}%
    \label{fig:robustness}%
\end{figure}

\begin{figure}[thpb]
  \centering
  \includegraphics[scale=0.3]{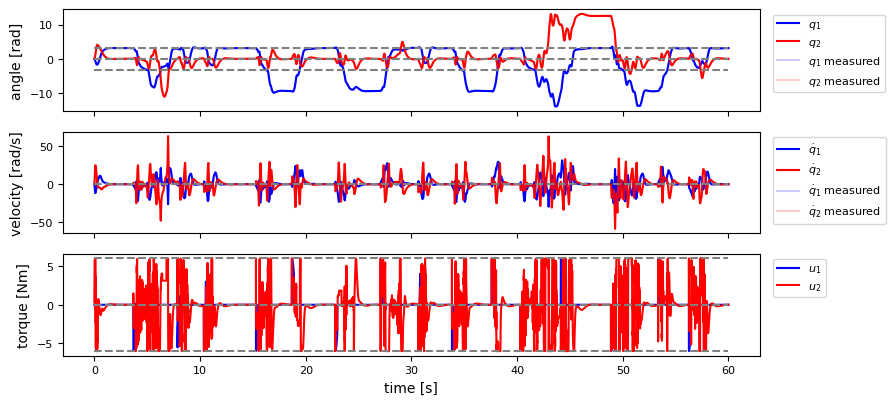}
  \caption{Simulation of the disturbed Acrobot system, controlled by the MPC}
  \label{fig:acrobot_traj}
\end{figure}

In the Acrobot setup the controller achieved a performance score of 0.508 with a total Uptime of 30.484 seconds. The robustness metrics are shown in Fig. \ref{fig:robustness}. The biggest loss in performance comes from the torque noise which is significantly less robust with Acrobot. Similar to what was observed with the Pendubot, the Acrobot is particularly weak to changes in model parameters related to the first, thus the inactive joint. It also reemphasizes the importance of finding a good strategy to deal with the Friction.
In many experiments the \lstinline|SQP_RTI| led to significant performance issues, only in the most recent experiments that could be overcome by fine tuning, allowing for a speed up in computation.
Overall the controller needed longer to re-stabilize after disturbances occurred but could always swing the pendulum back up in reasonable time as can be seen from Fig.\ref{fig:acrobot_traj}. Interestingly, it can be seen that disturbances are handles differently than in case of the Pendubot. The Acrobot accelerates into the direction of the disturbance but does not stop at the lower fixed point $[\pi, \pi]$. It continues until completing a full rotation, swings through the top fixed point $[0,0]$ with enough energy to complete a second rotation but now accelerating into the oposite direction to reduce the speed and finally stabilize at the top.  

\begin{table}[h]
\caption{Parameters used with Acrobot/Pendubot}
\label{table:options}
\begin{center}
\begin{tabular}{ |c|c| }
\hline
Attribute& Value\\
\hline
\lstinline|Nlp_max_iter|& \lstinline|40|\\
\lstinline|fallback_on_solver_fail|& \lstinline|true|\\
\lstinline|hpipbm_mode|& \lstinline|ROBUST|\\
\lstinline|max_solve_time|& \lstinline|0.01|\\
\lstinline|mpc_cycle_dt|& \lstinline|0.001|\\
\lstinline|prediction_horizon|& \lstinline|0.5|\\
\lstinline|qp_solver|& \lstinline|PARTIAL_CONDENSING_HPIPM|\\
\lstinline|qp_solver_tolerance|& \lstinline|0.001|\\
\lstinline|solver_type|& \lstinline|SQP_RTI|\\
\lstinline|warm_start|& \lstinline|true|\\
\hline
\end{tabular}
\end{center}
\end{table}

\begin{table*}[h]
\caption{Controller Options}
\label{table:classParameters}
\begin{center}
\begin{tabular}{ |c|c|c|p{6cm}| }
\hline
Attribute& Type& Default& Description\\
\hline
\lstinline|N_horizon|& \lstinline|float|& \lstinline|20|& number of shooting nodes\\
\lstinline|prediction_horizon|& \lstinline|float|& \lstinline|0.5|& prediction horizon\\
\lstinline|Nlp_max_iter|& \lstinline|int|& \lstinline|500|& maximum number of NLP iterations\\
\lstinline|max_solve_time|& \lstinline|float|& \lstinline|1.0|& Maximum time before solver timeout\\
\lstinline|solver_type,|& \lstinline|string|& \lstinline|SQP_RTI|& \lstinline|in ("SQP", "DDP", "SQP-RTI")|\\
\lstinline|wrap_angle|& \lstinline|bool|& \lstinline|0.5|& wether or not angles bigger than 360 deg are translated to $\theta \mod 360$ \\
\lstinline|warm_start|& \lstinline|bool|& \lstinline|True|& solver does some initial iterations to find a good initial guess\\
\lstinline|scaling|& \lstinline|int[]|& \lstinline|np.full(N_horizon, 1)|& scaling for the cost on nodes 1-N\\
\lstinline|nonuniform_grid|& \lstinline|bool|& \lstinline|False|& Timesteps $t_n$ are growing in size with there distance from $t_0$\\
\lstinline|use_energy_for_terminal_cost|& \lstinline|bool|& \lstinline|False|& wether in the terminal cost the energy is used instead of the state x\\
\lstinline|fallback_on_solver_fail|& \lstinline|bool|& \lstinline|False|& uses next $x$ of stored old solution if the NLP is not feasible\\
\lstinline|friction_compensation_on_inactive_joint|& \lstinline|float|& \lstinline|0.5|& inactive joint is set to be capable to exert a torque of 0.5 Nm as friction compensation\\
\lstinline|mpc_cycle_dt|& \lstinline|float|& \lstinline|0.01|& \lstinline|frequency of the mpc|\\
\lstinline|pd_tracking|& \lstinline|bool|& \lstinline|False|& \lstinline|use PID Controller|\\
\lstinline|outer_cycle_dt|& \lstinline|float|& \lstinline|0.001|& \lstinline|timestep of the integrated PID controller|\\
\lstinline|pd_KP|& \lstinline|float|& \lstinline|None|& \lstinline|Gain for position error for the PID Controller|\\
\lstinline|pd_KD|& \lstinline|float|& \lstinline|None|& \lstinline|Gain for integrated error for the PID Controller|\\
\lstinline|pd_KI|& \lstinline|float|& \lstinline|None|& \lstinline|N_horizon|\\
\hline
\end{tabular}
\end{center}
\end{table*}
\addtolength{\textheight}{-12cm}  

\section*{ACKNOWLEDGMENT}
Many thanks to Jonathan Frey and Armin Nurkanović from university of Freiburg for their developments on acados and their helpful discussions and inputs on tuning the controller.

\end{document}